\documentstyle[preprint,eqsecnum,aps]{revtex}
\begin{document}
\draft
\title{
\rightline{\rm \normalsize KIAS-P98006}
\vspace*{-0.4truecm}
\rightline{\rm \normalsize INFNCA-TH-9803}
NEUTRINO SELF-ENERGY AND DISPERSION IN A MEDIUM WITH
MAGNETIC FIELD}
\author{A. Erdas \thanks{erdas@vaxca2.unica.it}}
\address{
Dipartimento di Fisica dell' Universit\`a di Cagliari
and I.N.F.N. Sezione di Cagliari,
\\
Cittadella Universitaria, S.P. per Sestu Km 0.700,
I-09042 Monserrato (CA), Italy
}
\author{C. W. Kim \thanks{kim@eta.pha.jhu.edu}}
\address{
Department of Physics and Astronomy,
The Johns Hopkins University, Baltimore, MD 21218 \thanks{permanent address}
\\ School of Physics, 
Korea Institute for Advanced Study, Seoul 130-012, Korea}
\author{T. H. Lee 
\thanks{thlee@physics.soongsil.ac.kr}}
\address{
Department of Physics,
Soong Sil University, Seoul 156-743, Korea}
\maketitle
\vspace*{-0.4truecm}
\begin {abstract} 
We calculate the one-loop thermal 
self-energy of a neutrino in a constant 
and homogeneous magnetic field
to all orders in the magnetic field strength using Schwinger's 
proper time method. We obtain the dispersion relation under various conditions.
\end {abstract}
\section{Introduction}
The study of neutrino properties in a medium in the presence of a magnetic 
field is very important in the context of neutrino physics and of
a possible solution to the solar neutrino puzzle, and it has been
widely studied in the literature. It has been shown that, in the
weak-field limit, the presence of a magnetic field in the medium
modifies the neutrino index of refraction \cite{dolivo}, but,
for massless left-handed neutrinos the interaction with the magnetic field 
does neither flip chirality nor induce flavor transitions.
In a more recent paper \cite{elmfors2} the dispersion relation of neutrinos
in magnetized media is investigated and the authors calculate
the local part of the neutrino self-energy in a CP-asymmetric medium
to all orders in the magnetic field strength 
$B$ and also the term linear in $B$ for the
self energy in a CP-symmetric medium at temperatures much higher than
the electron mass $m$ and much lower than the $W$-mass $M$.
In this work we calculate the real part of the thermal self-energy
in a medium to all orders in the magnetic field, for all
temperatures below the critical temperature of the $SU(2)_L\otimes
U(1)$ standard model and for all momenta of the neutrino.
Our more general result reproduces all the results in the literature
\cite{dolivo,elmfors2}. 

In Section II we calculate the thermal self-energy to all orders in $eB$
using the real-time method of finite temperature field theory. 
In Section III, after showing that our results for a CP-asymmetric 
plasma agree with
those of Refs. \cite{dolivo,elmfors2}, we derive a simple expression,
valid to all orders in the magnetic field, for the self-energy and
dispersion relation in the presence of a non-relativistic electron gas.
In Section IV we use our main result of Section II to derive a simple
expression for the self-energy in a CP-symmetric medium when
$m\ll T \ll M$ to all orders in $eB$, and compute the dispersion
relation. We conclude in Section V. 

\section{ Exact finite temperature self-energy in a constant
magnetic field}
In this Section we use the exact propagators in a constant,
homogeneous magnetic field ${\bf B}$ to calculate the real part of the 
neutrino thermal self-energy in a magnetized medium. We take the
magnetic field oriented along the $z$-direction and the metric
to be $g^{\mu \nu} = {\rm diag}(-1,+1,+1,+1)$. Using Schwinger's proper 
time method \cite{schwinger}, it is possible to obtain the exact
expression for the vacuum propagator of the charged lepton
$S_0(x',x'')$ \cite{schwinger,dittrich}, the $W$-boson 
$G_0^{\mu \nu}(x',x'')$ \cite{erdasfeld} and the charged scalar
$\Delta_0(x',x'')$ \cite{dittrich,erdasfeld}
\begin{eqnarray}
S_0(x',x'')&=&i\phi^\ast
(x',x'')\!\int{d^4k\over (2\pi)^4}e^{ik\cdot (x'-x'')}
\int_0^\infty \!\!{ds\over\cos eBs}\,
{\exp{\left[-is\left(m^2-i\epsilon+k^2_{\scriptscriptstyle{\parallel}}+
k^2_{\scriptscriptstyle{\perp}}{\tan eBs\over eBs}\right)\right]}}
\nonumber \\
&&\times
\left[(m-
\not\! k_{\scriptscriptstyle{\parallel}})e^{-ieBs\sigma_3}-
{\not\! k_{\scriptscriptstyle{\perp}}\over \cos eBs}
\right],
\label{2_01}
\end{eqnarray}
\begin{eqnarray}
G_0^{\mu \nu}(x',x'')&=&i\phi(x',x'')\int{d^4k\over (2\pi)^4}
e^{ik\cdot (x'-x'')}
\int_0^\infty \!\!{ds\over\cos eBs}\,
{\exp{\left[-is\left(k^2_{\scriptscriptstyle{\parallel}}+
k^2_{\scriptscriptstyle{\perp}}{\tan eBs\over eBs}\right)\right]}}
\nonumber \\
&&\times\left\{
e^{-is(M^2-i\epsilon)}[g^{\mu \nu}_{\scriptscriptstyle{\parallel}}
+(e^{2eFs})^{\mu\nu}_{{\scriptscriptstyle{\perp}}}]
+\left({e^{-is(M^2-i\epsilon)}-e^{-is(\xi M^2-i\epsilon)}\over M^2}\right)
\right.
\nonumber \\
&&\times
\left[\left(k^{\mu}+k_{\lambda}
F^{\mu \lambda}{\tan eBs\over B}\right)
\left(k^{\nu}+k_{\rho}
F^{\rho \nu}{\tan eBs\over B}\right)
\right.
\nonumber \\
&&\left.\left.+i{e\over 2}\left(F^{\mu\nu}-
g^{\mu\nu}_{{\scriptscriptstyle{\perp}}}B\tan eBs
\right)\right]
\right\},
\label{2_02}
\end{eqnarray}
\begin{equation}
\Delta_0(x',x'')=i\phi(x',x'')\int{d^4k\over (2\pi)^4}
e^{ik\cdot (x'-x'')}
\int_0^\infty\!\!{ds\over\cos eBs}\,
{\exp{\left[-is\left(\xi M^2-i\epsilon+k^2_{\scriptscriptstyle{\parallel}}+
k^2_{\scriptscriptstyle{\perp}}{\tan eBs\over eBs}\right)\right]}},
\label{2_03}
\end{equation}
where the subscript $0$ on the propagators indicates explicitly that
these are vacuum propagators,
$-e$ is the charge of the electron, $m$ the electron mass, 
$M$ the $W$-mass,
$\xi$ the gauge parameter,
$B=|{\bf B}|$, $F^{\mu \nu}$ is the electromagnetic field strength
tensor whose only non-vanishing components are $F^{12}=B=-F^{21}$.
Also in Eqs. (2.1), (2.2), and (2.3) we have
\begin{equation}
\phi(x',x'')=\exp\left(
i {e\over 2}x''_\mu F^{\mu \nu} x'_\nu
\right),
\label{2_04}
\end{equation}
\begin{equation}
\sigma_3=\sigma^{12}={i \over 2}[\gamma^1, \gamma^2],
\end{equation}
for any 4-vector $a^\mu$ we define 
$a^\mu_{\scriptscriptstyle{\parallel}}$ and
$a^\mu_{\scriptscriptstyle{\perp}}$ as
\begin{equation}
a^\mu_{\scriptscriptstyle{\parallel}}=(a^0,0,0,a^3)\;\;\;,\;\;\;
a^\mu_{\scriptscriptstyle{\perp}}=(0,a^1,a^2,0)
\end{equation}
and we have
\begin{equation}
\left(e^{2eFs}\right)^{{\mu}\nu}_{{\scriptscriptstyle{\perp}}}=
(g^{\mu\nu})_{\scriptscriptstyle{\perp}} \cos 2eBs 
+F^{\mu\nu} {\sin 2eBs\over B}\,\,.
\end{equation}
This notation follows closely the one in Refs. \cite{dittrich,erdasfeld}.
The real-time thermal propagators are easily constructed starting from
the proper-time form of the propagators in vacuum. For the fermion we write
the vacuum propagator as
\begin{equation}
S_0(x',x'')=\phi^\ast(x',x'')\!\int{d^4k\over (2\pi)^4}e^{ik\cdot (x'-x'')}
S_0(k),
\label{2_05}
\end{equation}
and obtain the following expression \cite{elmfors2,elmfors1}
of the propagator $S(x',x'')$ in the presence of an external magnetic field
and a plasma at non-zero temperature and density, 
\begin{equation}
S(x',x'')=\phi^\ast(x',x'')\!\int{d^4k\over (2\pi)^4}e^{ik\cdot (x'-x'')}
S(k),
\label{2_08}
\end{equation}
where the translationally invariant part $S(k)$
of the propagator in a medium is defined in terms of the translationally 
invariant part $S_0(k)$ of the vacuum propagator and of the fermion
occupation number $f_F(k^0)$ at temperature $T$ and chemical potential
$\mu$
\begin{equation}
S(k)=S_0(k)-f_F(k^0)\Bigl[S_0(k)-S_0^\ast (k)\Bigr]\,\,.
\label{2_09}
\end{equation}
Notice that in Eq. (\ref{2_09}) the piece proportional
to the occupation number represents the thermal part of the propagator,
which we call $S_T(k)$. The fermion occupation number is defined as 
\begin{equation}
f_F(k^0)=f_F^+(k^0)\theta(k^0)+f_F^-(k^0)\theta(-k^0)
\label{2_09a}
\end{equation}
with
\begin{equation}
f_F^\pm(k^0)={1\over e^{\pm(k^0-\mu)/T}+1}\,\,.
\end{equation}
The same procedure is employed to obtain the propagators in a
medium of the $W$-boson, $G^{\mu\nu}(x',x'')$, 
and of the charged scalar, $\Delta(x',x'')$:
\begin{equation}
G^{\mu\nu}(x',x'')=\phi(x',x'')\!\int{d^4k\over (2\pi)^4}e^{ik\cdot (x'-x'')}
G^{\mu\nu}(k)
\label{2_10}
\end{equation}
with
\begin{equation}
G^{\mu\nu}(k)=G^{\mu\nu}_0(k)+f_B(k^0)\Bigl[G^{\mu\nu}_0(k)-
{G^{\mu\nu}_0}^\ast(k)\Bigr]
\label{2_11}
\end{equation}
where the boson occupation number $f_B(k^0)$ is defined as
\begin{equation}
f_B(k^0)={1\over e^{|k^0|/T}-1}
\end{equation}
and
\begin{equation}
\Delta(x',x'')=\phi(x',x'')\!\int{d^4k\over (2\pi)^4}e^{ik\cdot (x'-x'')}
\Delta(k)
\label{2_12}
\end{equation}
with
\begin{equation}
\Delta(k)=\Delta_0(k)+f_B(k^0)\Bigl[\Delta_0(k)-\Delta_0^\ast (k)\Bigr]\,\,.
\label{2_13}
\end{equation}
The three Feynman diagrams shown in
Fig. 1 contribute to the neutrino self-energy in a medium in magnetic field.
The tadpole
diagram $\Sigma_{tad}$ gives only a local contribution
to the self-energy, while the bubble diagram with a $W$-boson 
$\Sigma_{bub}(x',x'')$ and the bubble diagram with
a scalar $\Sigma_{scal}(x',x'')$ contain also non-local terms. They
are given by
\begin{equation}
\Sigma_{tad}=-ig_Z^2{\rm tr}
\left[\gamma_\mu (c^f_V-c^f_A\gamma_5)S(x,x) \right]\gamma_\nu \gamma_L 
Z^{\mu\nu}(0),
\label{2_14}
\end{equation}
\begin{equation}
\Sigma_{bub}(x',x'')=i\left({g\over \sqrt{2}}\right)^2\gamma_R 
\gamma_\mu S(x',x'') \gamma_\nu \gamma_L G^{\mu\nu}(x',x''),
\label{2_19}
\end{equation}
\begin{equation}
\Sigma_{scal}(x',x'')=i\left({g\over \sqrt{2}}\right)^2\left({m\over M}
\right)^2\gamma_R S(x',x'')\gamma_L \Delta(x',x''),
\label{2_20}
\end{equation}
where $Z^{\mu\nu}(0)$ is the vacuum propagator of the neutral $Z$-boson
at zero momentum, 
\begin{equation}
g_Z={g\over 2\cos\theta_W}
\,\,,\gamma_R={1+\gamma_5\over2}\,\,\,,\,\,\,
\gamma_L={1-\gamma_5\over2},
\label{2_16}
\end{equation}
and the index $f$ runs over $e$, $p$, $n$, since we consider a medium where
electrons, protons and neutrons are present, and
\begin{equation}
c^e_V=-c^p_V=c_V\,\,,\,\,\,\,c^e_A=-c^p_A=c_A\,\,,\,\,\,\,c^n_V=c^n_A
=-{1\over 2}
\label{2_17}
\end{equation}
with
\begin{equation}
c_V=-{1\over 2}+2\sin^2\theta_W\,\,,\,\,\,\,c_A=-{1\over 2}\,\,.
\label{2_18}
\end{equation}

We compute the self-energy of an electron neutrino in a medium. (A generalization to
muon or tau neutrino is straightforward.) Using the expression of the 
fermion propagator from Eqs. (\ref{2_01}), (\ref{2_08}), and (\ref{2_09})
we obtain the following expression of the tadpole term in a 
medium with non-vanishing electron chemical potential 
\begin{eqnarray}
\Sigma_{tad}&=&4{g_Z^2\over M_Z^2}\int^{+\infty}_{-\infty}\!\!{dk \over
(2\pi)^4}k
f_F(k)\int^{+\infty}_{-\infty}\!\!ds e^{-[is(m^2-k^2)+|s|\epsilon]}
\left(
{\pi\over is}\right)^{3/2}(eBs)
\nonumber \\
&&\times\gamma_R\left[ c_V\gamma^0\cot eBs
-ic_A\gamma^3 \right],
\label{2_23}
\end{eqnarray}
where the term linear in $B$ has a negative sign because in this paper we 
take the electron charge to be $-e$. The generalization to the case of 
non-zero nucleon chemical potential is quite straightforward and will 
be discussed briefly in Sec. III.
After we change variable in the $k$-integration
and perform the $s$-integration, Eq. (\ref{2_23}) becomes \cite{elmfors2}
\begin{equation}
\Sigma_{tad}={g^2_Z\over
M_Z^2}\gamma_R\left[c_V\gamma^0(N_e-N_{\bar{e}})
-c_A \gamma^3 (N_e^0-N_{\bar{e}}^0)
\right]\gamma_L ,
\label{2_24}
\end{equation}
where the net number density of electrons in a magnetic field is 
defined as \cite{elmfors2}
\begin{equation}
N_e-N_{\bar{e}}={|eB| \over 2\pi^2}\int_0^{\infty}dk_z
\sum_{n=0}^{\infty}\sum_{\lambda=\pm1}\left[f_F^+(E_{n,\lambda,k_z})
-f_F^-(-E_{n,\lambda,k_z})\right]
\label{2_25}
\end{equation}
and the energies of the Landau levels are
\begin{equation}
E_{n,\lambda,k_z}^2=m^2+k^2_z+|eB|(2n+1-\lambda),
\label{2_26}
\end{equation}
where $n$ is the orbital quantum number and $\lambda$ the spin
quantum number. The quantity $N_e^0-N_{\bar{e}}^0$ in Eq.
(\ref{2_24}) represents the net number density in the lowest
Landau level, and is given by
\begin{equation}
N_e^0-N_{\bar{e}}^0={|eB| \over 2\pi^2}\int_0^{\infty}dk_z
\left[f_F^+(E_{0,1,k_z})-f_F^-(-E_{0,1,k_z})\right]\,\,.
\label{2_27}
\end{equation}

Starting from Eqs. (\ref{2_19}) and (\ref{2_20}) we obtain the expressions
of the two bubble contributions to the self-energy, using the boson and
scalar propagators of Eqs. (\ref{2_02})-(\ref{2_03}) and
(\ref{2_10})-(\ref{2_13}).
We do the calculation in the Feynman gauge, and take $\xi=1$.
The two particles in the loop are oppositely charged, and therefore
the phase factors $\phi(x',x'')$ cancel, yielding a result which is
explicitly translationally invariant
\begin{equation}
\Sigma_{bub}(x',x'')=\int{d^4p\over (2\pi)^4}e^{ip\cdot (x'-x'')}
\Sigma_{bub}(p).
\label{2_28}
\end{equation}
The bubble contribution to the self-energy contains a vacuum part 
and two thermal parts due to contributions of 
thermal fermions, which in this case are electrons, and thermal bosons respectively
\begin{equation}
\Sigma_{bub}(p)=\Sigma_{bub}^0(p)+
\Sigma_{bub}^F(p)+\Sigma_{bub}^B(p)\,\,.
\label{2_29}
\end{equation}
The scalar contribution $\Sigma_{scal}$ is also translationally invariant 
and contains three pieces as in Eq. (\ref{2_29}).
For the expression of the vacuum part of $\Sigma_{bub}$
we refer the reader to Ref. \cite{erdasfeld},
here we are interested in the thermal parts which, in the 
$\xi=1$ gauge, are given by
\begin{eqnarray}
\Sigma_{bub}^F(p)&=&i{g^2\over 2}\gamma_R\gamma_\mu \int {d^4k\over
(2\pi)^4}f_F(k^0)\int_{-\infty}^{+\infty}ds
{e^{-[is(m^2+\tilde{k}^2)+|s|\epsilon]} \over \cos eBs}
\left[(m-
\not\! k_{\scriptscriptstyle{\parallel}})e^{-ieBs\sigma_3}-
{\not\! k_{\scriptscriptstyle{\perp}}\over \cos eBs}
\right]
\nonumber \\
&&\times\int_0^{\infty}dt
{e^{-it(M^2+\tilde{q}^2-i\epsilon)} \over \cos eBt}
[g^{\mu \nu}_{\scriptscriptstyle{\parallel}}
+(e^{2eFt})^{\mu\nu}_{{\scriptscriptstyle{\perp}}}]\gamma_\nu\gamma_L ,
\label{2_30}
\end{eqnarray}
\begin{eqnarray}
\Sigma_{bub}^B(p)&=&-i{g^2\over 2}\gamma_R\gamma_\mu \int {d^4k\over
(2\pi)^4}f_B(k^0)\int_{0}^{\infty}ds
{e^{-is(m^2+\tilde{q}^2-i\epsilon)} \over \cos eBs}
\left[(m-
\not\! q_{\scriptscriptstyle{\parallel}})e^{-ieBs\sigma_3}-
{\not\! q_{\scriptscriptstyle{\perp}}\over \cos eBs}
\right]
\nonumber \\
&&\times\int_{-\infty}^{+\infty}dt
{e^{-[it(M^2+\tilde{k}^2)+|t|\epsilon]} \over \cos eBt}
[g^{\mu \nu}_{\scriptscriptstyle{\parallel}}
+(e^{2eFt})^{\mu\nu}_{{\scriptscriptstyle{\perp}}}]\gamma_\nu\gamma_L ,
\label{2_31}
\end{eqnarray}
where $q=p-k$,
and we use the shortened notation
\begin{equation}
s\tilde{k}^2=sk^2_{\scriptscriptstyle{\parallel}}+
k^2_{\scriptscriptstyle{\perp}}{\tan eBs\over eB}\;,\;\;\;\;\;
\;\;\;\;\;\;\;
t\tilde{k}^2=tk^2_{\scriptscriptstyle{\parallel}}+
k^2_{\scriptscriptstyle{\perp}}{\tan eBt\over eB}\,\,,\,\,
\label{2_32}
\end{equation}
and the same for $\tilde{q}$.
The $\vec{k}$-integration is done after a suitable change of the 
$\vec{k}$-variable, to obtain 
\begin{eqnarray}
\Sigma_{bub}^F(p)&=&-{g^2\over 2}\int^{+\infty}_{-\infty}
{dk^0\over (2\pi)^3}
f_F(k^0)\int_0^\infty ds \int_{-1}^{+1}du\left({\pi\over is}\right)^{1/2}
{eBs\over
\sin eBs}\exp\left\{-is\left[um^2+(1-u)M^2\right.\right.
\nonumber \\
&&\left.\left.-u\left(k^0\right)^2-(1-u)\left(p^0-k^0\right)^2+u(1-u)
\left(p^3\right)^2+{\sin eBsu\sin eBs(1-u)\over
eBs \sin eBs}p^2_{{\scriptscriptstyle{\perp}}}\right]\right\}
\nonumber \\
&&\times
\gamma_R\left[e^{-i\sigma_3
eBs(2-u)}\left(\gamma^0k_0+(1-u)\gamma^3p_3\right)+
{\sin eBs(1-u)\over \sin eBs}\not\!p_{\scriptscriptstyle{\perp}}\right]
\gamma_L ,
\label{2_34}
\end{eqnarray}
\begin{eqnarray}
\Sigma_{bub}^B(p)&=&{g^2\over 2}\int^{+\infty}_{-\infty}{dk^0\over (2\pi)^3}
f_B(k^0)\int_0^\infty ds \int_{-1}^{+1}du\left({\pi\over is}\right)^{1/2}
{eBs\over
\sin eBs}\exp\left\{-is\left[(1-u)m^2+uM^2\right.\right.
\nonumber \\
&&\left.\left.-u\left(k^0\right)^2-(1-u)\left(p^0-k^0\right)^2+u(1-u)
\left(p^3\right)^2+{\sin eBsu\sin eBs(1-u)\over
eBs \sin eBs}p^2_{{\scriptscriptstyle{\perp}}}\right]\right\}
\nonumber \\
&&\times
\gamma_R\left[e^{-i\sigma_3
eBs(1+u)}\left(\gamma^0p_0-\gamma^0k_0+u\gamma^3p_3\right)+
{\sin eBsu\over \sin eBs}\not\!p_{\scriptscriptstyle{\perp}}\right]
\gamma_L .
\label{2_35}
\end{eqnarray}
Notice also that we have changed the variables of the $s$- and $t$-integrations
in the following way: $s\rightarrow su$ , $t\rightarrow s(1-u)$ 
in the fermionic part
and $s \rightarrow s(1-u)$ , $t \rightarrow su$ in the bosonic part.

The scalar contributions $\Sigma_{scal}^F$ and
$\Sigma_{scal}^B$ are easily obtained by replacing $g^2$ with
${g^2m^2\over 2M^2}$ in Eqs. (\ref{2_34}), (\ref{2_35}) and
$e^{-i\sigma_3 eBs (2-u)}$ with $e^{-i\sigma_3 eBs u}$ in Eq.
(\ref{2_34}) and $e^{-i\sigma_3 eBs (1+u)}$ with $e^{-i\sigma_3
eBs (1-u)}$ in Eq. (\ref{2_35}), respectively.

The expressions in Eqs. (\ref{2_34}) and (\ref{2_35}) are some of the
main results of our paper, which give the exact non-local
thermal self-energy for all values of the magnetic field, all values of 
the temperature that are below the critical temperature of $250$ GeV, all values
of electron chemical potential, and all values of the neutrino $4$-momentum.

\section{$CP$-asymmetric medium}

We consider now a $CP$-asymmetric medium at temperature $T\ll M$ 
and a magnetic field $B \ll M^2/e$. Under these conditions we 
can neglect thermal corrections to the $W$-propagator and 
simply use the vacuum propagator in the contact 
approximation
\begin{equation}
G^{\mu\nu}(x',x'')\simeq G_0^{\mu\nu}(x',x'')\simeq\phi(x',x'')\!
\int{d^4k\over (2\pi)^4}e^{ik\cdot (x'-x'')}{g^{\mu\nu}\over M^2}.
\label{3_1}
\end{equation}
To leading order in $1/M^2$ the scalar contribution is also negligible.

Using the Fierz-type identity \cite{dolivo}
\begin{equation}
\gamma_R\gamma^\mu A \gamma_\mu\gamma_L=-\gamma_R\gamma^\mu
{\rm Tr}\,[\gamma_\mu\gamma_L A],
\label{3_2}
\end{equation}
it is evident that $\Sigma^F_{bub}$ in the contact approximation is 
obtained from $\Sigma_{tad}$ of Eq. (\ref{2_14}) by making the
following replacements
\begin{equation}
{g^2_Z\over M^2_Z}\rightarrow {g^2\over 2M^2}\,\,\,,\,\,c_V
\rightarrow {1\over 2}\,\,\,,\,\,c_A\rightarrow {1\over 2}\,\,.
\label{3_3}
\end{equation}
Let us consider the thermal self-energy of an electron neutrino in an electron-rich
medium, to leading order in $1/M^2$. The sum of the tadpole and bubble
diagrams in the contact approximation is readily obtained making use
of Eq.(\ref{2_23}) and of the replacements mentioned in Eq. (\ref{3_3})
\begin{eqnarray}
\Sigma&=&{g^2\over M^2}\int^{+\infty}_{-\infty}\!\!{dk \over (2\pi)^4}k
f_F(k)\int^{+\infty}_{-\infty}\!\!ds e^{-[is(m^2-k^2)+|s|\epsilon]}
\left(
{\pi\over is}\right)^{3/2}(eBs)
\nonumber \\
&&\times\gamma_R\left[ (1+c_V)\gamma^0\cot eBs
-i(1+c_A)\gamma^3 \right]\gamma_L ,
\label{3_4}
\end{eqnarray}
where we have also used $M_Z=M/\cos\theta_W$.

In a weak-field limit we can write the self-energy as
\begin{equation}
\Sigma={g^2\over 4M^2}\gamma_R\left[(1+c_V)\gamma^0(n_e-n_{\bar{e}})
-(1+c_A) \gamma^3 (N_e^0-N_{\bar{e}}^0)
\right]\gamma_L+O(B^2),
\label{3_5}
\end{equation}
where the net number density of electrons in the absence of a magnetic field is
\begin{equation}
n_e-n_{\bar{e}}=2\int {d^3k\over
(2\pi)^3}[f_F^+(\omega_k)-f_F^-(-\omega_k)]
\label{3_6}
\end{equation}
with $\omega_k=\sqrt{|{\vec k}|^2+m^2}$, and $N_e^0-N_{\bar{e}}^0$ is 
the net number density of electrons in the lowest Landau level given by
Eq. (\ref{2_27}). The quantity $\gamma^3 (N_e^0-N_{\bar{e}}^0)$ can
be written as
$$
-2e\!\not\!\!B \int {d^3k\over (2\pi)^3}{1\over 2\omega_k}{d\over d\omega_k}
[f_F^+(\omega_k)-f_F^-(-\omega_k)]
$$
and therefore our results for the weak-field limit are in agreement
with those found in Ref.\cite{dolivo}. Note that the authors of Ref. [1] use the 
conventions opposite to ours $g^{\mu\nu}={\rm diag}
(+1,-1,-1,-1)$ and $e<0$ for the electron.

Let us now consider a nonrelativistic and nondegenerate limit
of the electron gas. Using the expansion of the cotangent as a power series
involving the Bernoulli numbers $A_l$, we rewrite Eq. (\ref{3_4}) as
\begin{eqnarray}
\Sigma&=&{g^2\over 4M^2}\gamma_R\left\{(1+c_V)\gamma^0
\left[1-\sum_{l=1}^{\infty}{(-1)^l\over(2l)!}|A_{2l}|(2eB)^{2l}
\left({\partial\over\partial m^2}\right)^{2l}
\right]
+(1+c_A)e\!\not\!\!B{\partial\over\partial m^2}
\right\}\gamma_L 
\nonumber \\
&&\times2\int {d^3k\over (2\pi)^3}[f_F^+(\omega_k)-f_F^-(-\omega_k)].
\label{3_7}
\end{eqnarray}
For a nonrelativistic and nondegenerate electron gas \cite{dolivo}
$$
f_F^+\simeq e^{-\beta(\omega_k-\mu)}\,\,\,\,\,{\rm and}\,\,\,\,f_F^-=0
$$
with the electron energy $\omega_k\simeq m+|{\vec k}|^2/2m$, and therefore
$$
{\partial f_F^+\over\partial m^2}\simeq-{\beta\over 2m}f_F^+.
$$
So for the nonrelativistic electron gas Eq. (\ref{3_7})
yields
\begin{equation}
\Sigma=\gamma_R(b\!\not \! u+c\not\!\!B)\gamma_L
\label{3_8}
\end{equation}
with
\begin{equation}
b=-{g^2\over 4M^2}(1+c_V)n_e{\mu_B B\over T}\coth \left({\mu_B B\over
T}\right)
\,\,\,\,\,\,\,\,\,{\rm and}\,\,\,\,\,\,\,\,\,c=-{g^2\over 4M^2}(1+c_A)n_e{\mu_B \over
T} ,
\label{3_9}
\end{equation}
where $\mu_B=e/2m$ is the Bohr magneton, $n_e$ is
the number density of electrons as defined in Eq. (\ref{3_6}),
$u^\mu$ is the 4-velocity of the medium and $\not \! u=
-\gamma^0$ in the reference frame of the medium.
This result is exact to all orders in the magnetic field 
provided that $T<< m$ and $\mu_B B<\pi T$.
The dispersion relation for electron neutrino in a nonrelativistic 
electron gas
is obtained by setting $(\not\!p+b\!\not \! u+c\not\!\!B)^2=0$, which, 
to lowest order in $1/M^2$, yields
\begin{equation}
E=|{\vec p}|+c{\hat p}\cdot{\vec B}-b,
\label{3_10}
\end{equation}
where the neutrino $4$-momentum is $p^\mu=(E,{\vec p})$.
The dispersion relation of Eq. (\ref{3_10}) shows that
electron neutrinos in the presence of a magnetic field 
and in a nonrelativistic electron gas have an anisotropic index of refraction $n$
\begin{equation}
n=1+{g^2\over 4M^2}n_e\left({\mu_B B\over T}\right)
{(1+c_A){\hat p}\cdot{\hat B}-(1+c_V)
\coth \left({\mu_B B\over T}\right)\over |{\vec p}|}
\label{3_11}
\end{equation}
and an effective mass $m_{\nu_e} = -b$. 
This $m_{\nu_e}$ is calculated from the one-loop diagram at finite
temperature in a magnetic field and is exact to all orders in the 
magnetic field strength for $\mu_B B<\pi T$.
This mass is similar in magnitude in the limit $\mu_B B/T\rightarrow 0$
as the well-known effective potential $V_C =\sqrt{2}G_F n_e$ induced 
by the coherent scatterings of a electron 
neutrino off electrons \cite{Wolfen}.

Finally we briefly consider the self-energy of an electron neutrino in
a medium where electrons, neutrons and protons are all present. 
Our results in the contact approximation agree with those of 
Ref. \cite{elmfors2}, and we write the self-energy exact to all orders in $B$
as
\begin{eqnarray}
\Sigma&=&{g^2\over 4M^2}\gamma_R\left\{\left[(1+c_V)(N_e-N_{\bar{e}})
-c_V(N_p-N_{\bar{p}})-{1\over 2}(N_n-N_{\bar{n}})\right]\gamma^0
\right.
\nonumber \\
&&\left. -(1+c_A) (N_e^0-N_{\bar{e}}^0)\gamma^3
\right\}\gamma_L ,
\label{3_12}
\end{eqnarray}
where $N_f-N_{\bar{f}}$ is the net number density of fermions in a magnetic field
as 
defined in Eq. (\ref{2_25}) and $ N_e^0-N_{\bar{e}}^0$ is the net number
density of 
electrons 
in the lowest Landau level, as defined in Eq. (\ref{2_27}).
In this Section we are considering temperatures much smaller 
than the typical nucleon mass 
and fields that are not extremely strong, therefore 
the nucleon magnetization is suppressed 
by their heavier mass relative to the electrons
\cite{elmfors2}.
For this reason we do not have a term proportional to
$\gamma^3$ times the nucleon density in Eq. (\ref{3_12}).

The dispersion relation for an electron neutrino is obtained immediately
\cite{elmfors2}
\begin{equation}
E=-b+|{\vec p}+{\vec c}|
\label{3_13}
\end{equation}
with
\begin{equation}
b=-{g^2\over 4M^2}\left[(1+c_V)(N_e-N_{\bar{e}})
-c_V(N_p-N_{\bar{p}})-{1\over 2}(N_n-N_{\bar{n}})\right]
\label{3_14}
\end{equation}
and
\begin{equation}
{\vec c}=
-{g^2\over 4M^2}(1+c_A) (N_e^0-N_{\bar{e}}^0){\hat B}.
\label{3_15}
\end{equation}

\section{$CP$-symmetric medium}

In a medium that is charge-symmetric, the local contributions to the
self-energy vanish, and we only need to consider the two bubble diagrams, but 
only the one with $W$-boson is significant because the scalar diagram is
suppressed by a factor of $m^2/M^2$.

In the weak-field approximation and in the $\xi=1$ gauge, we obtain the
following expression for the thermal part of the neutrino self-energy,
valid for any temperature below the critical temperature of the 
$SU(2)_L\otimes U(1)$ model, and for all values of the neutrino $4$-momentum
\begin{eqnarray}
\Sigma^F&=&{g^2\over 16\pi^2}\int_0^\infty k^2dk
\left({\partial\over\partial m^2}+2{\partial\over\partial M^2}\right)
f_F(\omega_k)\gamma_R
\left\{{e\not\!\!B\over k|{\vec p}|}L^-_1(k)\right.
\nonumber \\
&&\left.
+{e\gamma^0{\vec B}\cdot{\vec p}\over
\omega_k|{\vec p}|^2}
\left[4-{M^2-m^2-E^2+|{\vec p}|^2\over 2k|{\vec p}|}
L_1^+(k)-
{E\omega_k\over k|{\vec p}|}
L_1^-(k)\right]
\right\}\gamma_L ,
\label{4_1}
\end{eqnarray}
\begin{eqnarray}
\Sigma^B&=&{g^2\over 16\pi^2}\int_0^\infty k^2dk
\left({\partial\over\partial m^2}+2{\partial\over\partial M^2}\right)
f_B(\Omega_k)\gamma_R
\left\{{eB\sigma_3\not\!p_{\scriptscriptstyle{\parallel}}
\over \Omega_k k|{\vec p}|}
L_2^+(k)+{e\not\!\!B\over k|{\vec p}|}L_2^-(k)\right.
\nonumber \\
&&
\left.+{e\gamma^0{\vec B}\cdot{\vec p}\over
\Omega_k|{\vec p}|^2}
\left[4+{m^2-M^2-E^2+|{\vec p}|^2\over 2k|{\vec p}|}
L_2^+(k)-
{E\Omega_k\over k|{\vec p}|}
L_2^-(k)\right]
\right\}\gamma_L ,
\label{4_2}
\end{eqnarray}
where the neutrino $4$-momentum is $p^\mu=(E,{\vec p})$, $\omega_k=
\sqrt{m^2+k^2}$, $\Omega_k=\sqrt{M^2+k^2}$ and
the logarithmic functions \cite{weldon,quimbay} are
\begin{eqnarray}
L_1^\pm (k)&=&\ln\left({m^2-M^2+E^2-|{\vec p}|^2- 2E\omega_k
-2k|{\vec p}|\over m^2-M^2+E^2-|{\vec p}|^2- 2E\omega_k
+2k|{\vec p}|}\right)
\nonumber \\
&&
\pm \ln\left({m^2-M^2+E^2-|{\vec p}|^2+ 2E\omega_k
-2k|{\vec p}|\over m^2-M^2+E^2-|{\vec p}|^2+ 2E\omega_k
+2k|{\vec p}|}
\right),
\label{4_3}
\end{eqnarray}
\begin{eqnarray}
L_2^\pm (k)&=&\ln\left({M^2-m^2+E^2-|{\vec p}|^2+2E\Omega_k
+2k|{\vec p}|\over M^2-m^2+E^2-|{\vec p}|^2+ 2E\Omega_k
-2k|{\vec p}|}\right)
\nonumber \\
&&
\pm \ln\left({M^2-m^2+E^2-|{\vec p}|^2-2E\Omega_k
+2k|{\vec p}|\over M^2-m^2+E^2-|{\vec p}|^2- 2E\Omega_k
-2k|{\vec p}|}
\right)\,\,.
\label{4_4}
\end{eqnarray}
$\Sigma^F$ represents the contribution of thermal fermions (electrons) and
$\Sigma^B$ the contribution of thermal bosons, which are relevant for
$T\sim M$. Notice that the term proportional to $
eB\sigma_3\!\not\!p_{\scriptscriptstyle{\parallel}}$ in Eq. (\ref{4_2})
can be neglected because, for Dirac neutrinos on the mass-shell,
$\sigma_3\!\not\!p_{\scriptscriptstyle{\parallel}}={1\over 2}
\not\!p\sigma_3=0$.

We consider now temperatures $T\ll M$. In this case the thermal bosons do not 
contribute, and we only need to consider the vacuum part of the $W$-propagator.
For temperatures much lower than the $W$-mass, we can take the 
vacuum $W$-propagator in the $\xi=1$ gauge to be
\begin{equation}
G_0^{\mu\nu}(x',x'')\simeq\phi(x',x'')\!\int{d^4k\over (2\pi)^4}
e^{ik\cdot (x'-x'')}{g^{\mu\nu}\over M^2}\left(1-{k^2\over M^2}\right)
+O(1/M^6).
\label{4_5}
\end{equation}
The local piece does not contribute in a charge-symmetric medium, thus we
can write the self-energy in a magnetic field to order $1/M^4$ as
\begin{eqnarray}
\Sigma&=&-{g^2\over M^4}\int^{+\infty}_{-\infty}\!\!{dk \over (2\pi)^4}
f_F(k)\int^{+\infty}_{-\infty}\!\!ds e^{-[is(m^2-k^2)+|s|\epsilon]}
\left(
{\pi\over is}\right)^{3/2}(eB)
\nonumber \\
&&\times\gamma_R\left[ (i\sigma_3+\cot eBs)\left(2\gamma^0Esk^2
-i\gamma^3 p^3\right)-i{eBs\over \sin^2eBs} 
\not\!p_{\scriptscriptstyle{\perp}}\right]\gamma_L\,.
\label{4_6}
\end{eqnarray}
We now use the following identities
\begin{equation}
\cot eBs=i\sum_{n=0}^\infty\sum_{\lambda=\pm 1}e^{-ieBs(2n+1-\lambda)}
\,\,\,,\,\,\,
{eB\over\sin^2eBs}=-\sum_{n=0}^\infty\sum_{\lambda=\pm 1}
eB(2n+1-\lambda)e^{-ieBs(2n+1-\lambda)},
\label{4_7}
\end{equation}
change the variable of the $k$-integration, and obtain
\begin{eqnarray}
\Sigma&=&-{g^2\over M^4}{eB\over 2\pi^2}\int^{\infty}_0\!\!{dk_z}
\gamma_R \left[{f_F(E_{0,1,k_z})\over E_{0,1,k_z}}
\left(\gamma^0{\vec p}\cdot {\hat B}k^2_z+
{\vec \gamma}\cdot {\hat B}EE_{0,1,k_z}^2\right)\right.
\nonumber \\
&&\left.-\sum_{n=0}^\infty\sum_{\lambda=\pm 1}
{f_F(E_{n,\lambda,k_z})\over E_{n,\lambda,k_z}}
\left(\gamma^0EE^2_{n,\lambda,k_z}+
{\vec \gamma}\cdot{\hat B}{\vec p}\cdot {\hat B}k^2_z
+{eB\over 2}(2n+1-\lambda)
\not\!p_{\scriptscriptstyle{\perp}}
\right)\right]
\gamma_L ,
\label{4_8}
\end{eqnarray}
where $E$ and ${\vec p}$ are the neutrino energy and momentum respectively, 
$E_{n,\lambda,k_z}$ is the energy of a thermal electron in the
$n$-th Landau level (\ref{2_26}) and $E_{0,1,k_z}$ the energy
of thermal electrons in the lowest Landau level. 
The expression of the self-energy we derived is valid 
for $T\ll M$ and exact to all orders
in the magnetic field. It can be written in a more manifestly
covariant way using $\not\!p_{\scriptscriptstyle{\perp}}=
\not\!p+\gamma^0E-({\vec \gamma}
\cdot{\hat B})({\vec p}\cdot {\hat B})$.

In order to obtain the dispersion relation we need to add 
the contribution
of the bubble diagram with a $Z$-boson. This diagram is 
independent of the
magnetic field and has been computed several years ago \cite{notzold}.
In the $\xi=1$ gauge and in a $CP$-symmetric plasma it is
\begin{equation}
\Sigma_Z=-{7\pi^2g^2T^4\over 180M^4}\left(M^2\over M^2_Z\right)
\gamma_R\left(\gamma^0E+{1\over 4}\not\!p
\right)\gamma_L .
\label{4_9}
\end{equation}
We now concentrate on the situation where $m\ll T\ll M$ 
and $B\leq T^2 /e$. In the early
universe these conditions are reasonable between the $QCD$ phase
transition and nucleosynthesis. For the temperature and magnetic field
in this range we can perform the integration in Eq. (\ref{4_8})
and obtain
\begin{eqnarray}
\Sigma &=&-{g^2\over M^4}\gamma_R\left[{7\pi^2\over 90}T^4
\left(1+{M^2\over 2M^2_Z}\right)
\left(\gamma^0E+{1\over 4}\not\!p
\right)-{eT^2\over 24}(\gamma^0 {\vec p}\cdot {\vec B}+E
\not\!\!B)\right.
\nonumber \\
&&\left.-
{e^2\over 24\pi^2}\ln\left({m\over T}\right)
(-2B^2\gamma^0E+2{\vec p}\cdot{\vec B}\not\!\!B-B^2\not\!p)
\right]\gamma_L \,.
\label{4_10}
\end{eqnarray}
The part independent of $B$ agrees with the results of Ref. \cite{notzold}
and the part linear in $B$ agrees with the result of Ref. \cite{elmfors2}.
We were also able to obtain the $B^2$ term which could be important because of 
its dependence on $\ln\left({m\over T}\right)$. The resulting dispersion relation is
\begin{equation}
E =|{\vec p}|\left[1-{7\pi^2\over 90}{g^2T^4\over M^4}
\left(1+{M^2\over 2M^2_Z}\right)\right]
+{g^2T^2\over 12M^4}e{\vec B}\cdot{\vec p} 
+{g^2(eB)^2\over 12\pi^2M^4}|{\vec p}|\left[1-({\hat B}\cdot{\hat p})^2
\right]\ln\left({T\over m}\right) ,
\label{4_11}
\end{equation}
where the first two terms reproduce the result of Ref. 
\cite{elmfors2}, and the last one is very significant for $eB\sim T^2\gg 
m^2$.
\section{Conclusions}

We have studied the self-energy and dispersion relation of massless Dirac
neutrinos in the presence of a constant magnetic field in a plasma.
We obtain the self-energy in the $\xi=1$ gauge exact to all orders in the 
magnetic field, for all values of temperature below the critical temperature
($T_C$) of the Weinberg-Salam model and all values of electron chemical
potential. We then consider a $CP$-asymmetric plasma at temperature
$T\ll M$. We have shown that in the weak-field limit our result agrees with
those in the literature \cite{dolivo,elmfors2} and then derive
the self-energy and dispersion relation to all orders in the $B$-field for an 
electron neutrino in a nonrelativistic and nondegenerate electron plasma,
obtaining the index of refraction and effective mass of neutrino.
The effective mass $m_{\nu_e} =-b$ given in Eq. (3.9) is calculated
from the one-loop diagram at finite temperature in a constant and 
homogeneous magnetic field to all orders in the magnetic field strength.
Its actual magnitude in the limit $\mu_B B /T \rightarrow 0$ is similar
to that of the effective potential energy $V_C =\sqrt{2} G_F n_e$ induced
by the coherent scatterings of a electron neutrino off electrons. 
\cite{Wolfen}

Finally, we consider a charge-symmetric plasma, and derive from our
main result of Section II an expression for the self-energy in a
weak-field limit valid for $0\le T\le T_C$. This expression could be 
very useful for studying magnetic properties of neutrinos
for $T\sim M$. We then obtain the self-energy to all orders in $B$
for $T\ll M$ and then derive a very simple formula for the self-energy
for $m\ll T\ll M$ and $B\le T^2/e$ that contains 
also the term quadratic in $B$, 
while the terms linear and independent of $B$ are shown to agree with
results from the literature \cite{elmfors2,notzold}. We also obtain
the dispersion relation.
\acknowledgements
We wish to thank Gordon Feldman for helpful discussions.
A. Erdas wishes to thank the High Energy Theory Group of 
the Johns Hopkins University for the hospitality extended to him
during his several visits.
This work is supported in part by the
Basic Science Research Institute Program, Ministry of
Education, Korea, 1997, Project No. BSRI-97-2418 and in part by
Soong Sil University(THL).

\begin{figure}
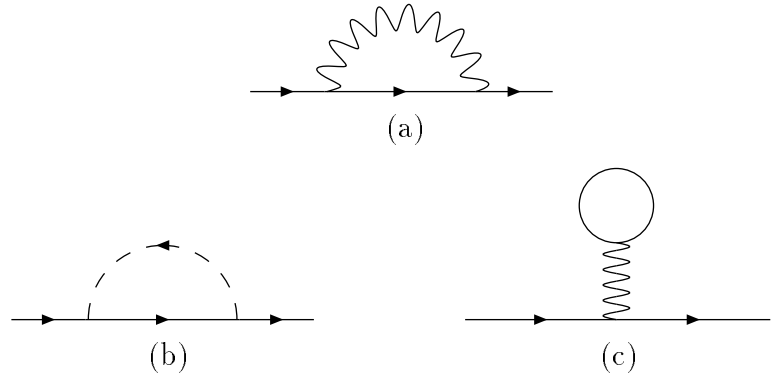

\caption{
The 3 diagrams relevant to the one-loop self-energy calculation; 
(a) the tadpole,
(b) the bubble diagram with $W$-boson and (c) the bubble diagram 
with scalar.}
\end{figure}
\end{document}